\documentclass[aps,prl,twocolumn,showpacs]{revtex4}	
\usepackage{graphicx,graphics,color}
\usepackage{mathtools} 
\usepackage{amssymb}

\begin{document}

\today

\title{
Judging Model Reduction of Chaotic Systems via Optimal Shadowing Criteria
}

\author{Jie Sun}
\email{sunj@clarkson.edu}
\author{Erik M. Bollt}
\email{bolltem@clarkson.edu}
\author{Takashi Nishikawa}
\email{tnishika@clarkson.edu}
\affiliation{Department of Mathematics \& Computer Science, Clarkson University, Potsdam, NY 13699-5815}

\begin{abstract}
A common goal in the study of high dimensional and complex system is to model the system by a low order representation.
In this letter we propose a general approach for assessing the quality of a reduced order model for high dimensional chaotic systems. The key of this approach is the use of optimal shadowing, combined with dimensionality reduction techniques. Rather than quantify the quality of a model based on the quality of predictions, which can be irrelevant for chaotic systems since even excellent models can do poorly, we suggest that a good model should allow shadowing by modeled data for long times; this principle leads directly to an optimal shadowing criterion of model reduction. This approach overcomes the usual difficulties encountered by traditional methods which either compare systems of the same size by normed-distance in the functional space, or measure how close an orbit generated by a model is to the observed data. 
Examples include interval arithmetic computations to validate the optimal shadowing.
\end{abstract}

\maketitle

Model reduction is an important concept found across science and engineering. Approximating gross scale features of high dimensional systems is a fundamental question which occupies a great deal of time and energy in the study of such disparate mathematical fields as PDE theory, time-delay systems, networked dynamical systems, and where-ever high dimensional problems naturally arise from the underlying science from which come the models. The POD method for example~\cite{holmes1998turbulence} is a popular way to produce a basis set for high-dimemensional data from solutons of PDEs, onto which the resulting Galerkin projections are optimal in the sense of a fastest decaying time-average power spectrum.
Underlying such techniques there is usually the common thread of minimization of the $\ell_2$ distance in the functional space between the actual system and its reduced order model - models are considered best in the Banach space. However, for chaotic systems,  use of the $\ell_2$ minimization criteria to compare the two functions for determining whether a model is good  may not be relevant, since two functions can be close in an underlying Banach space, but exhibit dramatically different dynamical properties~\cite{skufca2008concept}.  

\begin{figure}[ht]
\centering
\includegraphics*[width=0.45\textwidth]{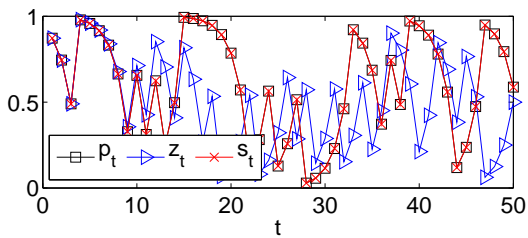}
\caption{
Illustration of the difficulty in judging a model by comparing orbits directly. 
In the upper panel, a noisy numerical orbit $\mathbf{p}=\{p_{t}\}_{t=1}^{50}$ of the logistic map is shown (in black square). This orbit satisfies $p_{t+1}=4{p_t}(1-p_t)+\delta_t$ where $\delta_t$ is uniformly distributed in $[-2^{10},2^{10}]$. Blue triangle represents a true, noiseless orbit $\mathbf{z}=\{z_{t}\}_{t=1}^{50}$ with $z_1=p_1=0.8724{...}$. Note that although $\mathbf{z}$ is close to $\mathbf{p}$ for initial times, after about $10$ steps they start to diverge. On the other hand, starting with $s_1=0.8723{...}$, we found a true orbit $\mathbf{s}=\{s_t\}_{t=1}^{50}$, shown in (red) crosses, which is able to match the entire noisy orbit $\mathbf{p}$. Although generated by the same model, $\mathbf{p}$ and $\mathbf{s}$ apparently leads to different conclusions about the model quality if we were to judge a model by comparing time series naively.
}
\label{1_LogisticNoisyOrbit}
\end{figure}

Likewise, a reasonable model, even a perfect model, may quickly produce quickly and dramatically different simulation results - it is well known that comparing time-series from simulations is an unworkable criterion of model comparison due to sensitive dependence.
When random noise or modeling error is introduced, as is arguably always the case in practice, even a seemingly perfect model would suffer from conflicting judgements. 
The sensitivity to perturbations prevents us from the comparing chaotic systems by direct comparison of their trajectories, since even (almost) identical systems would fail such a measure of comparison. See Fig.~\ref{1_LogisticNoisyOrbit} as an example.

To judge a model reduction, it is too much to hope that a model will be capable to reproduce trajectories of the full system, due to the chaotic nature of the system, as well as technical details of comparing trajectories which arise from systems of different dimensionality. We assert that such comparisons are meaningless good or bad because the expectation that the results will always be bad. Instead, we will judge a model to be a good representation if its trajectories can numerically \textit{shadow} trajectories of the full system. In this sense, the model is producing plausible solutions, if not the actual simulations.

Shadowing was introduced initially to rigorously verify the existence of a true orbit from a model to a computer generated orbit which is usually noisy~\cite{ANOSOV_67,BOWEN_75,HAMMEL_BAMS88,GREBOGI_PRL90,PALMER_BOOK}. Given a noisy orbit $\mathbf{p}=\{p_t\}$, a model generated orbit $\mathbf{x}=\{x_t\}$ is said to $\epsilon-$\textit{shadow} $\mathbf{p}$ if 
$||\mathbf{x}-\mathbf{p}||_{\infty}\equiv\sup_{t}||x_t-p_t||_{2}<\epsilon$
~\footnote{Here the choice of the first norm is crucial that it emphasizes the maximum possible difference between the two orbits, while the choice of the second norm is arbitrary.}.
From now on these subscripts of norms will be omitted unless otherwise specified. 

In terms of judging the model quality, we wish to associate the capability of the model to shadow observation with its quality. However, most shadowing techniques were developed only to find an arbitrary $\epsilon-$shadowing orbit, which may be far from optimal (there may be another shadowing orbit with a much smaller $\epsilon$), preventing us from a good judgement of the model. To overcome this ambiguity, we ask: what is the \textit{best} orbit the model can produce, to match the observed orbit? This amounts to
judge the quality of a model $f:D\rightarrow{D}$ for given observation $\mathbf{p}$ by the \textit{optimal shadowing distance}:
\begin{equation}
	\epsilon_{opt}\equiv\inf_{x_1\in{D}}||\mathbf{x}-\mathbf{p}||,
\end{equation}
where $D\subset\mathbb{R}^{m}$ and $\mathbf{x}$ is an orbit of $x_1$ under $f$~\footnote{We have specialized in discrete systems, while this concept can be naturally extended to continuous systems.}. For deterministic systems, this question is equivalent to finding the initial point which leads to a true orbit that can \textit{step-by-step} match the noisy orbit best.

Based on the concept of optimal shadowing, we focus on the question of how to understand the quality of a model reduction, meaning how well does a model of lower dimensionality represent the dynamics of the full system.  

\begin{figure}[ht]
\centering
\includegraphics*[width=0.45\textwidth]{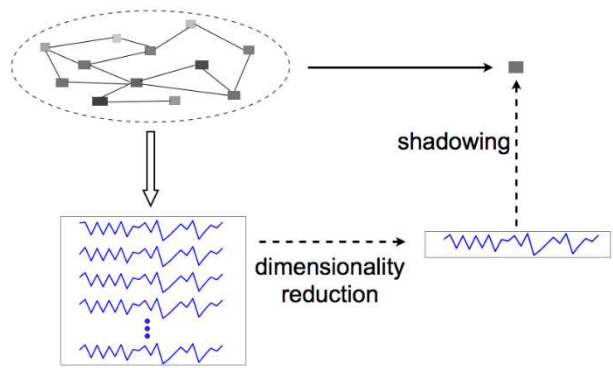}
\caption{General model reduction design cycle. (Down) A large scale system gives rise to a many variate time-series. (Bottom Across) Averaging across scales gives rise to a lower dimensional system with correspondingly fewer measurable variables in the output time-series. (Up) The step of judging model quality is usually overlooked in the design cycle.  To infer model quality, in some way the model must be required to remind of the full system.  Here we advocate that prediction is inappropriate due to sensitive dependence to initial conditions.  Instead we suggest an optimal shadowing criterion.     
}
\label{2_JudgingModelReduction}
\end{figure}

Since a high dimensional system and its reduced order model necessarily generate time series of different dimensions, there is currently no direct way of comparing two such models. Our approach to solve this problem can be illustrated by the diagram in Fig.~\ref{2_JudgingModelReduction}. Given a high dimensional system and its candidate reduced order model: we first generate time series from the original system; next, dimensionality reduction is performed to extract a low dimensional representation of the time series; finally, we look for an optimal shadowing orbit from the reduced order model to match the low dimensional time series. The reduced order model, being a simplification of the original one, suffers from two types of inexactness. The first type comes from dimensionality reduction, which accounts for the loss of information in simplifying the observation; the second type comes from shadowing, and is crucial for assessing the model quality of chaotic systems, which here accounts for the capability of the given model to generate one orbit that matches the observed (low dimensional) time series.

This approach allows us to quantify the quality of a model reduction even for chaotic systems, which is not likely to be achieved by traditional methods. Furthermore, the flexibility in emphasizing in between dimensionality reduction and shadowing errors allows one to adjust the measure of model quality in different situations depending on specific applications.

To illustrate this perspective, we consider the problem of modeling a system of coupled chaotic oscillators. Coupled oscillators have been studied extensively as prototypical of complex systems~\cite{PIKOVSKY_BOOK,ARENAS_REVIEW08}, with promising applications ranging widely from the modeling of flocking behavior~\cite{VICSEK_PRL95}, to mathematical epidemiology where collective behavior leads to mean field model of disease dynamics~\cite{STROGATZ_1993}, to mention a few. 
In any of these settings where many coupled oscillators may arise, it is natural to average across spatial scales so that a model with just a few oscillators may be meant to represent the system, in the sense that an element of the model may represent many elements of the whole. In the much the same way as a community analysis of complex networks where the topology allows partitioning into groups~\cite{Newman_2006,Porter_2009}, in dynamical systems we assert that groups of oscillators may exist with similar behavior.  When a system is modeled by a large collection of coupled oscillators the natural question is how might the simplified low order model captures similar properties of the original high order system? 

We choose to illustrate our approach by a system of coupled quadratic maps, described by:
\begin{equation}\label{CoupledSystem}
	x^{(i)}_{t+1} = a^{(i)}x^{(i)}_t(1-x^{(i)}_t) - \sigma\sum_{j=1}^{n}l_{ij}a^{(j)}x^{(j)}_t(1-x^{(j)}_t),
\end{equation}
where $\{x^{(i)}\}_{i=1,...,n}$ represent a set of coupled oscillators, $x^{(i)}_{t}\in\Re^{d}$ is the state of oscillator $i$ at time $t$; each individual oscillator is driven by a discrete logistic dynamics with parameter $a^{(i)}$, which allows possible mismatch of parameters between different individual oscillators, which is usually the case for a physical setting;
the second term describes the effective coupling between different oscillators through a discrete Laplacian matrix 
$L=[l_{ij}]_{n\times{n}}$, where for each $i$, $\sum_{j=1}^{n}{l_{ij}}=0$; and $\sigma$ is the coupling strength. The coupling function has been chosen to have the same form of the individual dynamics, which corresponds to the situation where each oscillator receives a direct signal from the output of its neighbors. 

\begin{figure}[ht]
\centering
\includegraphics*[width=0.3\textwidth]{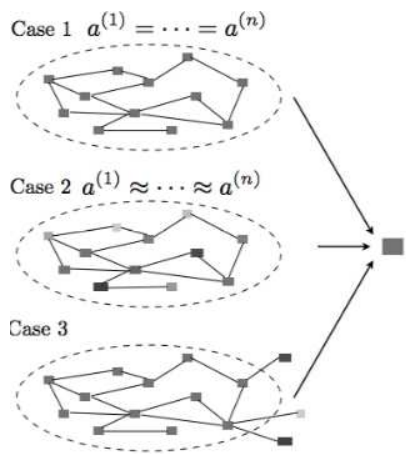}
\caption{Illustration of a model reduction of the coupled oscillator network. In the first case (top ellipse), all the oscillators are the same; in the second case (middle ellipse) the oscillators are mismatched; while in the third case, the network consists of a cluster of identical oscillators with a few outliers.
The rectangles represent individual oscillators (the width and color are used to highlight the difference of individual oscillators). In all cases, we are interested in whether model reduction is possible from the original system to a low dimensional system (represented by a single oscillator on the right).
}
\label{3_OSN}
\end{figure}

For this high ($n\times{d}$) dimensional coupled system, several questions are of particular interest, as initial exploration for the general problem, and will be answered in this letter. Fig.~\ref{3_OSN} serves as an illustration.

\newcounter{L1count}
\begin{list}{\arabic{L1count}}{\leftmargin=0.85em}\usecounter{L1count}
\item In what sense can we model a coupled identical oscillator network by a single oscillator?
\vspace{-0.25cm}
\item In what sense can we model a coupled {\it non-identical} oscillator network by a single oscillator?
\vspace{-0.25cm}
\item In what sense can we model a {\it nearly synchronized cluster} by a single oscillator? 
\end{list}
For question $1$, a general criteria is whether the system synchronizes or not. When the oscillators synchronize,
$\lim_{t\rightarrow\infty}||x^{(i)}_{t}-x^{(j)}_{t}||\rightarrow{0}$ for $\forall{i,j}$.
After transient, all the oscillators evolve in the same way, and the second term in Eq.~(\ref{CoupledSystem}) disappears (there will be no error in dimensionality reduction or shadowing). Any single oscillator $i$ is governed by the same dynamics:
	$x^{(i)}_{t+1} = {a}x^{(i)}_{t}(1-x^{(i)}_{t}).$
Thus we can perfectly model the coupled system by a single, low dimensional system:
$s_{t+1} = {a}s_{t}(1-s_t)$.

Questions $2$ and $3$ are intriguing. In these cases, the oscillators are unable to completely synchronize, thus a single oscillator model may not exactly represent the true collective behavior of the coupled system. In particular, if one chooses the average trajectory as a low dimensional representation of the high dimensional time series, then this average variable is governed by
\begin{equation}\label{TrueAvg}
	\bar{a}_{t+1} = \frac{1}{n}\sum_{i=1}^{n}a^{(i)}x^{(i)}_t(1-x^{(i)}_t) 
	- \frac{\sigma}{n}\sum_{i, j=1}^{n}l_{ij}a^{(j)}x^{(j)}_t(1-x^{(j)}_t) ,
\end{equation}
which depends essentially on \textit{every} single oscillator, implying that the dimension of the system is as high as the original coupled system. Even in the situation where the oscillators are \textit{nearly synchronized}~\cite{SUN_EPL09}: 
$\limsup_{t}||x^{(i)}_{t}-\bar{x}_{t}||\approx{0}$,
if one were to use mean-field approximation,replacing $x^{(i)}_{t}$ with 
$\bar{x}_t$ and $a^{(i)}$ with $\bar{a}$, resulting in a model:
\begin{equation}\label{ApprAvg}
	\bar{a}_{t+1} = \bar{a}\bar{x}_{t}(1-\bar{x}_{t}),
\end{equation}
then at each step this model generates error (comparing to the actual average state) which comes from the heterogeneity of the individual dynamics, and its effect might be tremendous depending on how the heterogeneity distributes among the oscillators. Nevertheless, our approach overcomes the difficulty and provide a quantitative measure of the reduced order model.

We shall illustrate this for case $2$ by the use of optimal shadowing for the average trajectory. As a matter of example, we will construct a network of logistic oscillators whose individual parameters $a^{(i)}$ are drawn uniformly from $[3.9998,4]$ in order to emphasize oscillator mismatch. We couple those oscillators  through an Erd\H os-R\'enyi network \cite{erd?s1961strength} of $n=1000$ and $p=0.1$ (the probability that any two nodes are joined by an edge), with coupling strength $\sigma=0.0075$.

\begin{figure}[ht]
\centering
\includegraphics*[width=0.4\textwidth]{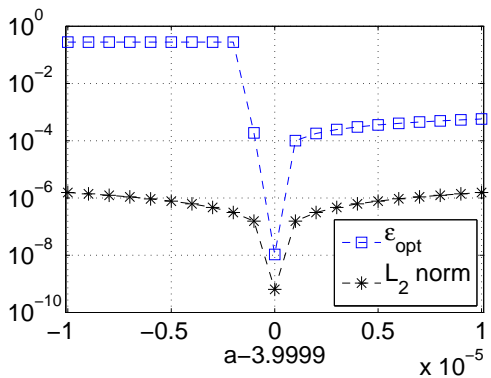}
\caption{(Color online)
Optimal shadowing distances of reduced order model for a coupled oscillator network. This system consists of mismatched logistic oscillators coupled through a network of $n=1000$ nodes and $p=0.1$, with coupling strength $\sigma=0.0075$. Blue square corresponds to the optimal shadowing distance for a one-parameter family of reduced order models $f(x)=ax(1-x)$; black star corresponds to the $\ell_2$ norm $\sqrt{\sum_{t=1}^{T-1}|f(x_t)-x_{t+1}|^2/T}$.
}
\label{4_ModelingViaShadowing}
\end{figure}

The dependence of optimal shadowing distance depends upon the parameter $a$ for a one-parameter family of reduced models $f(x)=ax(1-x)$. Here we use a finite trajectory of length $T=1000$ after transients. $\epsilon_{opt}$ are calculated by use of interval arithmetic, with the excellent package ``INTLAB" \cite{rump1999intlab}, in order to validate that we are representing reasonable upper bounds of the actual optimal shadowing distances.   Results are shown in Fig.~\ref{4_ModelingViaShadowing}, for a typical trajectory generated by the original network. It is interesting to note the difference between using the shadowing criteria in contrast to the  usual $\ell_2$ criteria: while the model error seems to depend symmetrically on $a$ under the $\ell_2$ criteria, shadowing is able to capture the asymmetry which seems to be more reasonable because of the increase of topological entropy for increasing $a$. Shadowing also has the advantage to judge how long the reduced order model is valid for the original system (the optimal shadowing distance increases non-smoothly when we take longer trajectories), another perspective the $\ell_2$ criteria does not provide. We have also obtained similar results in the case of modeling a nearly synchronized cluster (case $3$), which will be reported in a more detailed paper.

\begin{figure}[ht]
\includegraphics*[width=0.5\textwidth]{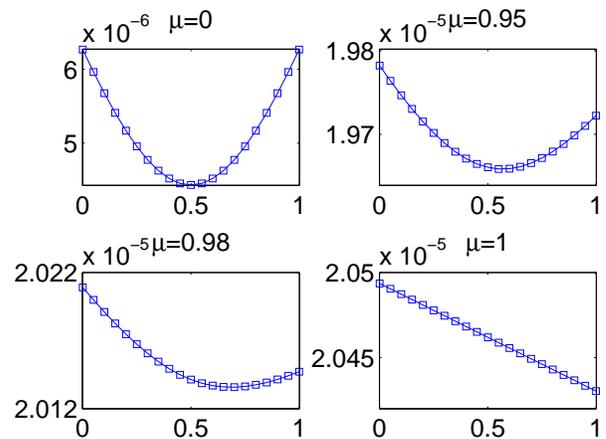}
\caption{(Color online)
Interplay between dimensionality reduction error and shadowing error. In each panel the horizontal axis corresponds to  $\lambda$ and vertical axis corresponds to the model reduction error: $\zeta(\lambda)=(1-\mu)\eta(\lambda)+\epsilon_{opt}(\lambda)$, for fixed $\mu$.
}
\label{5_InterplayDimensionShadowing}
\end{figure}

The above example demonstrates the judging of a model reduction by measurement of the optimal shadowing distance from a model to the average trajectory from the original system. Our choice to use such an average was selected to minimize the square distance to all other individual trajectories, i.e., the dimensionality reduction error.   To illustrate this perspective, consider a toy example where we have two logistic oscillators with parameters $3.9998$ and $4$, coupled through a network of Laplacian matrix $L=[2,-2;-1,1]$ with coupling strength $\sigma=0.25$. The dimensionality reduction of the time series $[\mathbf{x}^{(1)},\mathbf{x}^{(2)}]$ can be represented by a convex sum: 
$\mathbf{x}=(1-\lambda)\mathbf{x}^{(1)}+\lambda\mathbf{x}^{(2)}$. For given $\lambda$, the dimensionality reduction error can be defined as: $\eta(\lambda)=\sqrt{\sum_{i}||\mathbf{x}-\mathbf{x}^{(i)}||^2/(2T)}$ where $T$ is the length of $\mathbf{x}$.
In Fig.~\ref{5_InterplayDimensionShadowing} we show how one would obtain different dependence of the \textit{model reduction error}
$\zeta(\lambda)=(1-\mu)\eta(\lambda)+\epsilon_{opt}(\lambda)$ on $\lambda$. It is interesting to note especially in the last panel (lower right of Fig.~\ref{5_InterplayDimensionShadowing}) that when we emphasize purely on the modelability of the low dimensional system, then the trajectory from the single oscillator $\mathbf{x}^{(2)}$ would induce the best model (among the family of models $f(x)=ax(1-x)$). On the other hand, for other choice of $\mu$, the optimal $\lambda$ would change, not necessarily equals $1/2$, as expected. 

In general it will be interesting to ask such questions as in a large network, how shall we take the weighted average of individual trajectories to reach an optimal balance between dimensionality reduction and shadowing; or how nonlinear dimensionality reduction can be adopted in the case of generalized synchronization~\cite{SUN_SIADS09}. Some of the results will be reported in a future paper.

To summarize, we have proposed a general approach for assessing the quality of  reduced order models for high dimensional chaotic systems. The key in this approach is the unusual application of concepts from shadowing, toward the optimal shadowing criterion, combined with dimensionality reduction techniques. This approach overcomes perhaps overlooked problems inherent with  traditional methods of comparison which may either attempt to compare systems of the same size by measuring the distance in the functional space, or alternatively to measure how close an orbit generated by a model is to the observed data.  Both of these perspectives have fundamental flaws which our optimal shadowing based cost function overcomes.

{\it Acknowledgements ---}
E.M.B. and J.S. were supported by the Army Research Office under Grant 51950-MA. 

\bibliographystyle{apsrev}
\bibliography{JudgingModelViaShadowing_ref}


\end{document}